# Compact Modeling of pH-Sensitive FETs Based on Two-Dimensional Semiconductors

Tarek El Grour, Francisco Pasadas, Alberto Medina-Rull, Montassar Najari, Enrique G. Marin, Alejandro Toral-Lopez, Francisco G. Ruiz, Andrés Godoy, David Jiménez and Lassaad El-Mir

*Abstract*—We present a physics-based circuit-compatible model for pH-sensitive field-effect transistors based on two-dimensional (2D) materials. The electrostatics along the electrolyte-gated 2D-semiconductor stack is treated by solving the Poisson equation including the Site-Binding model and the Gouy-Chapman-Stern approach, while the carrier transport is described by the drift-diffusion theory. The proposed model is provided in an analytical form and then implemented in Verilog-A, making it compatible with standard technology computer-aided design tools employed for circuit simulation. The model is benchmarked against two experimental transition-metal-dichalcogenide (MoS$_2$ and ReS$_2$) based ion sensors, showing excellent agreement when predicting the drain current, threshold voltage shift, and current/voltage sensitivity measurements for different pH concentrations.

## I. INTRODUCTION

TWO-dimensional (2D) semiconductors combine excellent electrostatics and electronic transport properties with high surface-to-volume ratio, ultimate scaling limits, flexibility and the feasibility to large-scale processing and manufacturing, constituting great candidates for low-cost wearable and implantable biosensors [1], [2]. In this regard, field-effect transistors (FETs) based on transition metal dichalcogenides (TMDs) are receiving considerable attention due to their ability to perform label-free electrical detection of biological species such as proteins, DNA, and several bio-molecules [3]–[7]. Their use as ion-sensitive FETs (ISFETs) has also been widely demonstrated showing excellent detection capabilities and, in the particular case of pH, high stability and near-ideal pH voltage sensitivity, close to the Nernst limit (around 59 mV/ pH at room temperature) [4], [5], [8]–[10]. This state of the art at the device level foresees their soon application in advanced bio-electronics circuits. It is in this forthcoming arena where 2D-ISFETs compact modeling becomes extremely useful, not only to interpret electrical measurements or predict the device and circuit optimal operation, but also to assist and expedite the design of novel prototypes able to leverage the unique properties of TMDs, as well as to benchmark them against conventional technologies.

The work presented in this brief develops a comprehensive description of the 2D-ISFET operation by combining a verified compact model of 2D-semiconductor FETs (2DFETs) [11] with the modeling of the solid-liquid interface through the Site-Binding model and the Gouy-Chapman-Stern approach [12]. The compact model here proposed constitutes the first description of 2D-ISFETs for pH sensing that is compatible with standard technology computer-aided design (TCAD) tools employed for circuit simulation. In the following, we present the theory behind the electrostatics and carrier transport (Section II), and we validate the predictive capabilities of the model with measurements of two experimental TMD (MoS$_2$ and ReS$_2$) based ISFETs for pH sensing (Section III). Finally, the main conclusions are drawn in Section IV.

## II. DRAIN CURRENT MODELING OF 2D-ISFETS

Figure 1a shows a schematic depiction of a 2D-ISFET, whose structure consists of a FET where the top-gate metal is substituted by an electrolyte solution with a reference electrode immersed in it. In addition to the induced charge in the electrolyte, the electrostatic modulation of the carrier concentration in the 2D sheet is achieved via the reference electrode ($V_g$) and a bottom gate contact ($V_b$) coupled through a bottom dielectric. The top oxide acts as a barrier, guaranteeing

This work is supported in part by the Spanish Government under the projects TEC2017-89955-P, RTI2018-097876-B-C21 and PID2020-116518GB-I00 (MCIU/AEI/FEDER, UE); the FEDER/Junta de Andalucía under project B-RNM-375-UGR18; EC under Horizon 2020 projects WASP No. 825213 and GrapheneCore3 No. 881603. E.G. Marin gratefully acknowledges Juan de la Cierva Incorporación IJCI-2017-32297. A. Toral-Lopez acknowledges the FPU program (FPU16/04043). F. Pasadas acknowledges funding from PAIDI 2020 and Andalusian ESF OP 2014-2020 (20804). F. Pasadas and D. Jiménez also acknowledge the partial funding from the ERDF allocated to the Programa Operatiu FEDER de Catalunya 2014-2020, with the support of the Secretaria d'Universitats i Recerca del Departament d'Empresa i Coneixement of the Generalitat de Catalunya for emerging technology clusters to carry out valorization and transfer of research results. Reference of the GraphCAT project: 001-P-001702. (Corresponding authors: T. El Grour and F. Pasadas.)

T. El Grour and L. El-Mir are with LAPHYMNE Laboratory, Gabes University, Gabes, Tunisia (e-mail: grour_tarek@hotmail.fr).

F. Pasadas, A. Medina-Rull, E. G. Marin, A. Toral-López, F. G. Ruiz, and A. Godoy are with the PEARL Laboratory, Departamento de Electrónica y Tecnología de Computadores, Universidad de Granada, Granada 18071, Spain (e-mail: fpasadas@ugr.es).

M. Najari is with The Innovation and Entrepreneurship Centre, Jazan University, Jazan, Saudi Arabia.

D. Jiménez is with the Departament d'Enginyeria Electrònica, Escola d'Enginyeria, Universitat Autònoma de Barcelona, 08193 Bellaterra, Spain.

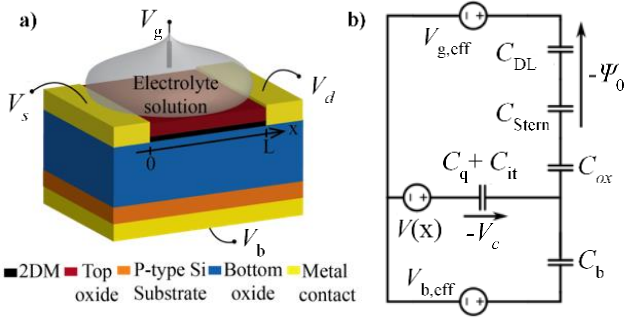

Fig. 1. a) Schematic depiction of a 2D-ISFET. b) Equivalent capacitive circuit of the 2D-ISFET.

an unambiguous field-effect transduction mechanism through the electrostatic control of the channel. The one-dimensional electrostatics along a vertical cut of the 2D-ISFET can be described using the equivalent capacitive circuit, shown in Fig. 1b, that sets the following charge-balance equation:

$$Q_{\text{net}}(x) + Q_{\text{it}}(x) = -C_{ox}\left(V_{g,\text{eff}} + \Psi_0 - V(x) + V_c(x)\right) - C_b\left(V_{b,\text{eff}} - V(x) + V_c(x)\right) \quad (1)$$

where $Q_{\text{net}}$ represents the overall net mobile sheet charge density at the 2D channel; $C_{ox} = \varepsilon_{ox}/t_{ox}$, ($C_b = \varepsilon_b/t_b$) is the top (bottom) oxide capacitance, with $\varepsilon_{ox}$ ($\varepsilon_b$) the top (bottom) dielectric constant and $t_{ox}$ ($t_b$) the top (bottom) oxide thickness. The top (bottom) overdrive voltage is $V_{g,\text{eff}} = V_g - V_{g0}$ ($V_{b,\text{eff}} = V_b - V_{b0}$), where $V_g$ ($V_b$) is the reference electrode (bottom gate) potential, and $V_{g0}$ ($V_{b0}$) comprises the work-function difference between the electrode (bottom gate) and the 2D channel as well as additional fixed charge due to impurities or doping [13]. Finally, $\Psi_0$ is the potential drop at the electrolyte-oxide interface (i.e. the surface potential).

The ISFET operation is grounded on the interaction between the surface charge density ($N_s$) [14] generated by the chemical reactions taking place at the top oxide-electrolyte interface, and the free charge carriers present in the semiconductor. This interaction, and specifically the adsorption of $H^+$ ions, is governed by the Site-Binding model [14], whose standard formulation [12] is adopted here. The pH variations of the electrolyte solution are related to $\Psi_0$ as

$$\Psi_0 = \ln[10]\left(\text{pH}_{\text{pzc}} - \text{pH}\right)\frac{\beta}{1+\beta}V_{\text{th}}$$
$$\text{pH}_{\text{pzc}} = \frac{\text{pK}_a + \text{pK}_b}{2} \quad (2)$$

where $V_{\text{th}} = k_BT/q$ is the thermal voltage, $k_B$ is the Boltzmann constant, $T$ is the temperature and $q$ is the elementary charge. pH$_{\text{pzc}}$ is the pH value at the charge neutrality point of the oxide surface, $\beta$ is defined by the Gouy–Chapman–Stern model and pK$_a$ and pK$_b$ correspond to the dissociation constants. At the electrolyte-insulator interface, the Gouy-Chapman approach describes the double layer as a function of the ionic strength in terms of the capacitance $C_{\text{DL}}$. However, this model overestimates the interface charge [12]. To correct this effect, a region depleted of ionic charges close to the surface is included, giving rise to the so-called Stern capacitance, $C_{\text{Stern}}$ [15]. Thus

$$\beta = qN_s\frac{\delta}{C_{\text{eq}}V_{\text{th}}}; \quad \delta = 2\cdot 10^{\frac{\text{pK}_a-\text{pK}_b}{2}}; \quad C_{\text{eq}} = \frac{C_{\text{DL}}C_{\text{Stern}}}{C_{\text{DL}}+C_{\text{Stern}}} \quad (3)$$

where $C_{\text{DL}}$ [12] can be computed as follows:

$$C_{\text{DL}} = \frac{\sqrt{8\varepsilon_W V_{\text{th}} q n_0}}{2V_{\text{th}}}; \quad n_0 = N_A i_0 \quad (4)$$

where $\varepsilon_W$ is the electrolyte dielectric permittivity; $n_0$ is the ionic charge concentration; $N_A$ is Avogadro constant (1/mol) and $i_0$ is the ionic molar concentration of the solution.

This way, the solid-liquid interface is modelled in (2) by a nonlinear voltage source ($\Psi_0$) which is dependent on pH, ionic concentration, the density of the surface ionizable sites, and their respective dissociation constants [16]. We combine such a dependent voltage source with an established compact model for 2DFETs [11]. In this regard, at the 2D semiconductor region, the main variables governing the carrier statistics, which are thoroughly explained in [11], are summarized as follows: i) the quasi-Fermi level ($-qV$), that must fulfill $V(x=L) = V_d$ ($V(x=0) = V_s$) at the drain (source) edge, where $L$ is the gate length and $V_d$ ($V_s$) is the drain (source) voltage; and ii) its shift with respect to the conduction (valence) band edge in an $n$-type ($p$-type) ISFET ($qV_c$). The source/drain electrodes are considered to be passivated, so that ions cannot be adsorbed by these metals. In addition, our model considers a trapped charge density in the semiconductor, $Q_{\text{it}}$, computed as [11], [17] $Q_{\text{it}} = qN_{\text{it}}/(1+\exp((V_c-V_{\text{it}})/V_{\text{th}}))$, where $N_{\text{it}}$ is the effective density of trap states and $-qV_{\text{it}}$ is the shift of the effective trap energy level respect to the conduction/valence band. Then, the trap capacitance $C_{\text{it}}$ can be calculated as [11]:

$$C_{\text{it}} = \frac{dQ_{\text{it}}}{dV_c} = \frac{qN_{\text{it}}}{2V_{\text{th}}}\frac{1}{1+\cosh[(V_c-V_{\text{it}})/V_{\text{th}}]} \quad (5)$$

We can, eventually, find an expression to evaluate the net sheet density assuming a parabolic dispersion relationship and incorporating Fermi–Dirac statistics [11]:

$$Q_{\text{net}}(x) = -q^2D_0V_{\text{th}}u(V_c); \quad u(V_c) = \ln\left[1+e^{-V_c/V_{\text{th}}}\right] \quad (6)$$

where $D_0 = g_K(m^K/2\pi\hbar^2)+g_Q(m^Q/2\pi\hbar^2)\exp[-\Delta E_2/k_BT]$ is the 2D density of states in a two-valley semiconductor, with $\hbar$ the reduced Planck's constant, $g_K$ ($g_Q$) the degeneracy factor and $m^K$ ($m^Q$) the band effective mass at the $K$ ($Q$) valley. In most of 2D TMDs the energy separation between the $K$ and $Q$ valleys, $\Delta E_2$, is only around $2k_BT$ [18], [19] and both valleys participate in the transport process. Further valleys are neglected because they are typically far away in energy to contribute to the electrical conduction under common bias conditions [20]. The quantum capacitance is given by $C_q = dQ_{\text{net}}/dV_c = C_{dq}(1-\exp(-u))$ [11], where $C_{dq} = q^2D_0$ is the degenerated quantum capacitance, i.e. the maximum value reachable when the 2D channel is highly degenerated ($V_c \ll -V_{\text{th}}$) [21].

As an explicit expression for $V_c$ as a function of the terminal biases cannot be obtained from (1) and (6), the strategy proposed in [22], [23] is considered. Basically, it implements a Verilog-A algorithm that iteratively evaluates the chemical potentials at the source and drain edges ($V_{cs} = V_c|_{V=V_s}$ and $V_{cd} = V_c|_{V=V_d}$, respectively), allowing the circuit simulator to solve such equations in a moderate run-time and, thus, achieving a circuit-compatible model.

Considering a drift-diffusion transport regime, the current of a 2D-ISFET can be accurately calculated as [11], [24]:

$$I_{ds} = \mu\frac{W}{L}C_{dq}V_{\text{th}}^2\left[\left(1+\frac{C_{dq}}{C_{t,eq}+C_b}\right)\left(\frac{u_s^2-u_d^2}{2}\right)+(e^{-u_d}-e^{-u_s})\right] \quad (7)$$

where $u_s = u(V_{cs})$ and $u_d = u(V_{cd})$; $W$ is the channel width; $\mu$ is the electron/hole mobility and $C_{t,eq}$ is the series combination of $C_{eq}$ and $C_{ox}$.

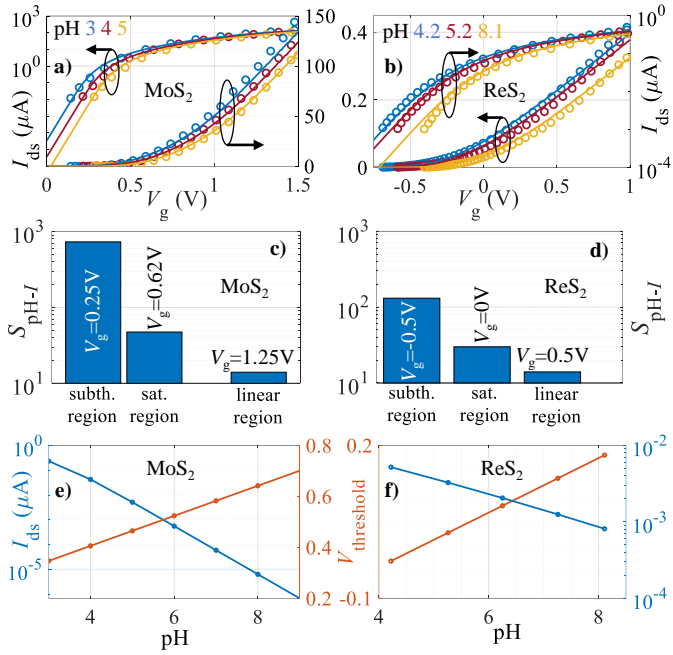

Fig. 2. Transfer characteristics in both logarithmic and linear scales of a) an *n*-type MoS$_2$-ISFET with 30-nm HfO$_2$ as top oxide and b) an *n*-type ReS$_2$-ISFET with a top oxide of 20-nm Al$_2$O$_3$; at different pH values. Symbols represent the measurements and solid lines the model outcome. The devices are reported in [4], [5] and described by the parameters collected in Table I. pH current sensitivity in different operation regions for c) the MoS$_2$-ISFET for a change of pH from 4 to 5 and d) the ReS$_2$-ISFET for a pH change from 4.2 to 5.2. Threshold voltage and drain current in the subthreshold region for e) the MoS$_2$-ISFET ($V_g$ = 0.25V, voltage sensitivity of 59 mV/pH) and f) the ReS$_2$-ISFET ($V_g$ = -0.5V, voltage sensitivity of 53.4 mV/pH).

The proposed analytical description is implemented in Verilog-A and included into Advanced Design System© (ADS).

### III. RESULTS AND DISCUSSION: THEORY VS. EXPERIMENTS

We have assessed the model against experimental measurements taken from a MoS$_2$-FET pH sensor by Sarkar *et al*. [4] (fabricated on 270 nm SiO$_2$/Si substrate and 30 nm of HfO$_2$ as gate dielectric) and a ReS$_2$-ISFET (fabricated on a 20 nm HfO$_2$/Si substrate with a top oxide of 20-nm thick Al$_2$O$_3$ that passivates the channel) reported by Liao *et al*. [5]. Table I summarizes the model parameters in both cases and the Site-Binding parameters ($N_s$, pK$_a$, pK$_b$) of both top oxide-electrolyte interfaces.

Figures 2a-b show our compact model results jointly with the measurements of the transfer characteristics ($I_{ds}$–$V_g$) of the MoS$_2$- and ReS$_2$-based devices, respectively, at three different pH values (pH = 3, 4, 5 for [4]; and pH = 4.22, 5.26, 8.11 for [5]) in both logarithmic and linear scales. It must be highlighted the very good agreement achieved in all regimes of operation, evidencing the predictive capabilities of the proposed model here proposed. A key figure of merit for pH sensors is the pH current sensitivity $S_{pH-I}$, defined as the relative change of the 2D-ISFET current corresponding to a unit change of the pH value, i.e., $S_{pH-I}$ = ($I_{pH2}$ − $I_{pH1}$) / $I_{pH1}$ ×100, where $I_{pH1}$ and $I_{pH2}$ are the currents at the two different pH values. $S_{pH-I}$ can be readily extracted from our model as shown in Figs. 2c-d. In the subthreshold region, the drain current has exponential dependence on the gate voltage reaching 731 (131) for the MoS$_2$-(ReS$_2$-)ISFET, while in saturation and linear regions the

TABLE I
2D-ISFET MODEL PARAMETERS

|  | MoS$_2$ | ReS$_2$ |  | MoS$_2$ | ReS$_2$ |
|---|---|---|---|---|---|
|  | SiO$_2$ | Al$_2$O$_3$ |  | SiO$_2$ | Al$_2$O$_3$ |
| $L$ (μm) | 5 | 1.72 | $g_K$ | 2 | 2 |
| $W$ (μm) | 20 | 3.01 | $g_Q$ | 6 | 6 |
| $t_{ox}$ (nm) | 30 | 20 | $V_{ds,ext}$ (V) | 1 | 0.1 |
| $\varepsilon_{ox}$ | 25 | 9 | $R_c$ (kΩ·μm) | 1 | 7.21 |
| $t_b$ (nm) | 270 | 20 | $\mu_0$ (cm$^2$/Vs) | 200 | 6.4 |
| $\varepsilon_b$ | 3.9 | 15.4 | $N_{it}$ (cm$^{-2}$) | 2.5·10$^{12}$ | 10$^{11}$ |
| $V_{g0}$ (V) | 0.59 | 0.175 | $V_{it}$ (eV) | 0.021 | 0.07 |
| $V_{b0}$ (V) | 0 | 0 | $\varepsilon_w$ | 80$\varepsilon_0$ | 80$\varepsilon_0$ |
| $m^K/m_0$ | 0.54 [29] | 1.53 [30] | pK$_a$ | 7 [31] | 10 [32] |
| $m^Q/m_0$ | 0.58 [29] | 1.53 [30] | pK$_b$ | 7 [31] | 6 [32] |
| $\Delta E_2$ (eV) | 0.07 [29] | 0.3 [33] | $i_0$ (mM) | 10 | 10 |
| $C_{Stern}$ (μF/cm$^2$) | 20 [34] | 20 [34] | $N_s$ (cm$^{-2}$) | 4·10$^{14}$ [31] | 8·10$^{14}$ [32] |

* $R_c$ represents the metal-2D semiconductor contact resistance, included by connecting lumped resistors to the drain and source terminals [35].

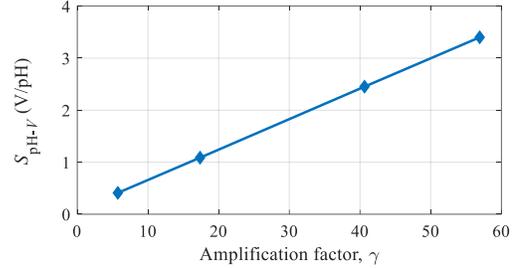

Fig. 3. Voltage sensitivity of the MoS$_2$-ISFET described in Table I, varying the bottom gate bias and setting $V_g$ = 1V for different amplification factors, γ = $C_{ox}$/$C_b$.

relationship becomes quadratic and linear [25], with values of 47 (30) and 14 (14), respectively. These predictions are in excellent agreement to those experimentally achieved: 713 (126), 53 (35) and 13 (14) in [4] ([5]). In addition, we have employed this model to determine the change of drain current as a function of the pH in the subthreshold region (Figs. 2e-f). The pH voltage sensitivity, $S_{pH-V}$ (i.e. the shift of threshold voltage due to pH), is found to be 59 mV/pH (53.4 mV/pH) for the MoS$_2$-(ReS$_2$-) ISFET, which is again in agreement with the measured pH sensitivity in [4] ([5]), a value close to the Nernst limit, showing the extraordinary potential of 2D-ISFETs for ion detection.

Furthermore, exploiting the model implementation a double-gate ISFET can be used to increase the threshold voltage shift by taking advantage of the electrostatic coupling between the two gate capacitances [26]. The change in the threshold voltage is in this case modified by the ratio $C_{ox}$/$C_b$ as follows:

$$\frac{\Delta V_{t,b}}{\Delta \text{pH}} = \left(\frac{C_{ox}}{C_b}\right)\frac{\Delta V_{t,g}}{\Delta \text{pH}} = \left(\frac{C_{ox}}{C_b}\right)\frac{\Delta \Psi_0}{\Delta \text{pH}} \qquad (8)$$

where $V_{t,g}$ ($V_{t,b}$) is the top (bottom) gate threshold voltage. Figure 3 shows $S_{pH-V}$ for $V_{t,b}$ considering a fixed $V_g$ = 1V, by applying the second-derivative method [27] to the transfer characteristics ($I_{ds}$–$V_b$) for different pH values of the MoS$_2$-ISFET. As shown in Fig. 3, $S_{pH-V}$ increases linearly with the pH sensitivity amplification factor, γ = $C_{ox}$/$C_b$ reaching values beyond the Nerst limit (~3.4 V/pH). This is consistent with the results achieved by Le *et al*. in [28], who demonstrated a measured $S_{pH-V}$ of 4.4 V/pH for γ = 33 for the particular case of monolayer MoS$_2$-based ISFETs.

## IV. Conclusion

A physics-based compact model for double-gate 2D semiconductor-based ion sensitive field-effect transistors has been developed and implemented in Verilog-A. It is therefore compatible with standard commercial circuit simulators. The code is available from corresponding authors upon reasonable request. The predictive capabilities have been benchmarked against experimental data from two devices based on different TMDs, bringing to light the robustness of the proposed approach to predict the electrochemical behavior of such devices. This model also allows a straightforward application to different 2D materials and represents a valuable contribution to speed up the design and testing of applications aiming the detection of analytes in chemical and biological experiments.